\documentclass[aps,prb,showpacs,superscriptaddress,twocolumn]{revtex4}
\usepackage{amsfonts}
\usepackage{txfonts}
\usepackage{epsfig}

\def\be{\begin{equation}} \def\ee{\end{equation}}
\def\bea{\begin{eqnarray}} \def\eea{\end{eqnarray}}

\def\nn{\nonumber}

\def\br{{\bf r}}

\begin{document}
\title{Strongly Correlated Topological Superconductors and Topological Phase Transitions via Green's Function}

\author{Zhong Wang}

\affiliation{
Institute for Advanced Study, Tsinghua University, Beijing,  China, 100084}

\author{Shou-Cheng Zhang$^{1,}$}

\affiliation{
Department of Physics, Stanford University, Stanford, CA 94305}

\begin{abstract}

We propose several topological order parameters expressed in terms of Green's function at zero frequency for topological superconductors, which generalizes the previous work for interacting insulators. The coefficient in topological field theory is expressed in terms of zero frequency Green's function. We also study topological phase transition beyond noninteracting limit in this zero frequency Green's function approach.

\end{abstract}

\maketitle

\section{Introduction}\label{sec:intro}

Recently topological phases have been among the central topics in condensed matter physics\cite{qi2010a, moore2010, hasan2010,qi2011}. These phases of matter are interesting in that their bulk is gapped but their surface is robustly gapless. Among these topological phases are topological insulators, including quantum Hall insulators\cite{laughlin1983}, and the recently discovered time reversal invariant topological insulators including the quantum spin Hall insulators\cite{kane2005b,bernevig2006c,konig2007} and its three-dimensional(3d) generalization\cite{qi2010a, moore2010, hasan2010,qi2011}. On the other hand, many superconductors have gapped fermionic spectrum in the bulk, therefore, they can also be topologically classified. For free fermion systems, the topological classification of insulators and superconductors are well established by previous works\cite{qi2008,schnyder2008,kitaev2009}. The central tools in these topological classifications are various band topological invariants\cite{kane2005b,fu2006,moore2007,fu2007b,qi2008,schnyder2008,wang2010a}, starting from the classical Thouless-Kohmoto- Nightingale-den Nijs(TKNN) invariant\cite{thouless1982}. In the presence of interactions, topological states of matter can be
generally defined in terms of the topological response functions within topological field theory\cite{qi2008}. Recently, there are great interests in the field of interacting topological insulators\cite{raghu2008,shitade2009,zhang2009b,seradjeh2009,pesin2010,fidkowski2010,
li2010,dzero2010,rachel2010,zhang2012,neupert2012,go2012}. Among the most urgent problems in the field of interacting topological insulators is to formulate topological invariants/topological order parameters to distinguish different topological phases. To this end, topological order parameters expressed in terms of Green's function for interacting topological insulators were proposed\cite{wang2010b}. There are much recent interest in the Green's function approach to interacting topological insulators\cite{wang2011,wang2011a,gurarie2011,chen2011,yoshida2011}. More recently, it was proposed that topological order parameters can be expressed in terms of Green's functions at zero frequency without losing any information\cite{wang2012,wang2012a}, which greatly simplifies numerical calculations\cite{go2012,wang2012b,budich2012} and analytical calculation.


The theory of interacting topological superconductors are still less investigated compared to topological insulators.
In Ref.\cite{wang2011b,ryu2012} it has been proposed that gravitational topological field theory can be formulated to describe interacting 3d time reversal symmetric topological superconductors\cite{schnyder2008,kitaev2009,qi2009b}, however, there is no explicit formula for the $\theta$ angle appearing in the $\theta$ term (even for noninteracting systems). Therefore, it is generally a difficult task to determine the topological class of a given interacting superconductor. Recently, there are several studies on interacting superconductors in 2d\cite{qi2012,ryu2012a,yao2012}
in which some judiciously chosen symmetries are imposed instead of the more conventional symmetries such as time reversal symmetry. In the present paper, we will focus only on the more conventional cases, namely superconductors without any additional symmetry and superconductors with time reversal symmetry, since these cases are closer to real material problems in experiments. By extending the previously mentioned Green's function approach to superconductors, we will obtain several simple yet general topological order parameters for interacting superconductors. These topological order parameters are expressed in terms of zero frequency Green's function, which are easy to compute by various numerical methods and analytical calculations. We also present simple examples in which interaction effect induce nontrivial topological phase transitions, which are nevertheless well described within our approach.

The rest of this paper is organized as follows. In Sec.\ref{sec:2d} we present topological order parameter for 2d chiral superconductors expressed in terms of zero frequency Green's function, and then in Sec.\ref{sec:application} we study a simple but nontrivial ansatz  to illustrate the application of our new formula in strongly correlated systems. In Sec.\ref{sec:3d} we present topological order parameter for 3d time reversal invariant topological superconductors. Then in Sec.\ref{sec:gravity} we obtain the gravitational $\theta$ angle in terms of Green's function, and show that it is exactly given by the winding number of zero frequency Green's function proposed in Sec.\ref{sec:3d}. In Sec.\ref{sec:gapless}, we discuss generalization of zero frequency Green's function approach to gapless systems. We then make several concluding remarks in Sec.\ref{sec:conclusion}, including finite size calculations and the general recipe of the zero frequency Green's function approach to topological insulators/supercondutors in all symmetry classes. Finally, in appendices we collect various exact Green's function identities for superconductors, which are useful in the main text, and detailed derivations of several equations in the main text.

\section{Correlated chiral topological superconductors in two dimensions}\label{sec:2d}

In this section, we will obtain topological order parameter for chiral superconductors. This topological order parameter is expressed in terms of zero frequency Green's function. This is analogous to the generalized first Chern number for quantum (anomalous) Hall insulators\cite{wang2012a}, however, there are differences between the two, among which the most prominent is the breaking of $U(1)$ gauge symmetry for superconductors. More precisely, the Green's function matrix contains four submatrices, which are nonetheless not independent with each other. The fermion operators are $c_{k\alpha},\,\alpha=1,2\dots N$, where $\alpha$ refers to any degree of freedom other than momentum. We can define the $2N$-component Nambu operator as $\Psi(k)= (c_k,c_{-k}^\dag)^T$. The Matsubara Green's function is defined as $\mathcal{G}_{\alpha\beta}(i\omega,k)=-\int_0^\beta e^{i\omega\tau}\langle T_\tau \Psi_\alpha(\tau,k)\Psi^\dag_\beta(0,k)$ with $\alpha,\beta=1,2\dots 2N$.  The Green's function $\mathcal{G}$ are decomposed into four submatrices as \bea \mathcal{G}(i\omega,k) = \left(
                                \begin{array}{cc}
                                G_A(i\omega,k) & G_B(i\omega,k) \\
                                G_C(i\omega,k) & G_D(i\omega,k) \\
                                \end{array}
                                \right) \eea
where \bea (G_A)_{\alpha\beta}(i\omega,k) &=&-\int_0^\beta d\tau e^{i\omega\tau}\langle T_\tau c_{k\alpha}(\tau,k)c_{k\beta}^\dag(0,k)\rangle \nn \\ (G_B)_{\alpha\beta}(i\omega,k)&=&-\int_0^\beta d\tau e^{i\omega\tau}\langle T_\tau c_{k\alpha}(\tau,k)c_{-k\beta}(0,k)\rangle \nn \\ (G_C)_{\alpha\beta}(i\omega,k)&=& -\int_0^\beta d\tau e^{i\omega\tau}\langle T_\tau c_{-k\alpha}^\dag(\tau,k)c_{k\beta}^\dag(0,k)\rangle \nn \\ (G_D)_{\alpha\beta}(i\omega,k)&=&-\int_0^\beta d\tau e^{i\omega\tau}\langle T_\tau c_{-k\alpha}^\dag(\tau,k)c_{-k\beta}(0,k)\rangle \eea

Following Ref.\cite{wang2012,wang2012a}, we diagonalize the inverse Green's function as \bea \mathcal{G}^{-1}(i\omega,k)|\alpha(i\omega,k)\rangle=\mu_\alpha(i\omega,k)|\alpha(i\omega,k)\rangle \label{eigen} \eea
We define those eigenvectors with $\mu_\alpha(0,k)>0$ as ``R-zero'', while those with $\mu_\alpha(0,k)<0$ as ``L-zero''\cite{wang2012,wang2012a}. From Eq.(\ref{hermitian}), we know that the Green's function is Hermitian at zero frequency, more explicitly, we have $\mathcal{G}^\dag(0,k) = \mathcal{G}(0,k)$, therefore, the eigenvalues of $\mathcal{G}(0,k)$ is real. The spaces spanned by R-zeros and L-zeros are named as R-space and L-space. From the property of Hermitian matrix we can see that vectors within R-space are orthogonal to those within L-space, therefore, we can define Berry connections and curvature in the R-space. With these preparations, we can now propose one of the central results of this paper, namely a generalized Chern number
\bea  C_1 = \frac{1}{2\pi} \int d^2k \mathcal{F}_{xy} \label{chern} \eea where $\mathcal{F}_{ij}=\partial_i \mathcal{A}_j - \partial_j \mathcal{A}_i$, and $\mathcal{A}_i = -i\sum_{R-space}\langle k\alpha |\partial_{k_i}|k\alpha\rangle$, where $|k\alpha\rangle$ are orthonormal vectors spanning the R-space. The simplest basis choice is $|k\alpha\rangle =|\alpha(i\omega=0,k)\rangle$. Eq.(\ref{chern}) strongly resembles the result for the quantum anomalous Hall insulators\cite{wang2012a}, however, the R-zeros in superconductor cases are mixtures of particle/hole components.
We also mention in passing that since only zero frequency Green's function is needed in Eq.(\ref{chern}), we can also use real frequency Green's function, which is the same as the Matsubara Green's function at zero frequency.  Since we only need zero frequency Green's function, Eq.(\ref{chern}) can be evaluated efficiently by various analytical and numerical methods.

Next we will show the relation of Eq.(\ref{chern}) to another topological invariant defined for chiral superconductors, namely
the winding number defined from interacting Green's function at all frequency, which is explicitly given as\cite{wang2010b,volovik2003} \bea N_2 =
\frac{1}{24\pi^2} \int d^2kd\omega {\rm Tr}[\epsilon^{\mu
\nu \rho} (\mathcal{G}\partial_\mu \mathcal{G}^{-1})(\mathcal{G}\partial_\nu \mathcal{G}^{-1})( \mathcal{G}\partial_\rho
\mathcal{G}^{-1})] \label{n2} \eea This topological quantum number measures the quantized thermal Hall effect\cite{read2000}. We can prove the equivalence \bea N_2=C_1 \label{derivation} \eea by similar calculation to its analogue in the quantum Hall effect\cite{wang2012a}. The key idea is to introduce a smooth deformation of $\mathcal{G}(i\omega,k)$ which does not change $N_2$. Due to intrinsic properties of Green's functions, a simple deformation connecting $\mathcal{G}(i\omega,k)$ to an ``effective noninteracting'' Green's function can be found (see Appendix \ref{sec:derivation} for details) , it can thus be shown that $N_2$ is fully determined by zero frequency Green's function. This leads exactly to $N_2=C_1$. The readers are referred to Appendix \ref{sec:derivation} for details of calculations, which is a straightforward generalization of the derivations in  Ref.\cite{wang2012a} .

Similar to the case\cite{wang2012a} of quantum Hall insulators,  Eq.(\ref{derivation}) shows that it is unnecessary to do frequency integral to obtain the topological invariant for superconductors, because zero frequency Green's function already contains sufficient information to determine the topological class. This greatly simplifies numerical and analytical calculations.

The topological invariant given in Eq.(\ref{chern}) is equivalent to the Chern number calculated from an effective ``noninteracting Bogoliubov-de Gennes Hamiltonian'' \bea h(k) =-\mathcal{G}^{-1}(\omega=0,k) \label{free} \eea therefore, we may be tempted to think that Eq.(\ref{chern}) is a generalization of the noninteracting topological invariants to the ``renormalized energy bands'', or ``Bogoliubov-de Gennes quasiparticles''. However, this picture is incorrect. Let us explain more about this important point following Ref.\cite{wang2012c}. In fact, in the ``renormalized bands'' or ``quasiparticle'' picture, we should use the self-consistent equation $\mathcal{G}^{-1}(\omega_\alpha,k)|u_\alpha(k)\rangle=0$, which reveals the approximate quasiparticle poles. Taking advantage of the Dyson equation $\mathcal{G}^{-1}(\omega_\alpha,k)=\omega_\alpha -h_0(k)-\Sigma(\omega_\alpha,k)$, we have \bea [h_0(k)+\Sigma(\omega_\alpha,k)]|u_\alpha(k)\rangle=\omega_\alpha |u_\alpha(k)\rangle \label{eigen-2} \eea where $h_0(k)$ is the free part of the Hamiltonian and $\Sigma(\omega,k)$ is the self energy generated by electron-electron interaction.  This ``renormalized bands'' or ``quasiparticle'' approach would suggest we define the topological invariant in terms of $|u_\alpha(k)\rangle$, however, this approach is different from the zero frequency Green's function approach.
Unlike this ``renormalized band'' picture, the zero frequency Green's function approach adopts a different equation\cite{wang2012c}  \bea [h_0(k)+\Sigma(0,k)]|k\alpha\rangle=\epsilon_\alpha |k\alpha\rangle \label{eigen-1} \eea  which follows from the Dyson equation $-\mathcal{G}^{-1}(0,k)=h_0(k)+\Sigma(0,k)$. Eq.(\ref{eigen-1}) is counterintuitive because the nonzero eigenvalue $\epsilon_\alpha$ suggests we self-consistently use $\Sigma(\epsilon_\alpha,k)$ instead of $\Sigma(0,k)$. As a comparison, we note that in Eq.(\ref{eigen-2}), $|u_\alpha(k)\rangle$ has the clear physical meaning as the `` quasiparticle''(or ``would-be quasiparticle'', because the concept of quasiparticle is most useful only when the lifetime is long), and the corresponding eigenvalues can be regarded as energy spectra. In contrary to Eq.(\ref{eigen-2}), in Eq.(\ref{eigen-1}) such clear physical meaning is absent for $|k\alpha\rangle$, and the corresponding eigenvalues $\epsilon_\alpha$ cannot be interpreted as energy spectra, because of the lack of self-consistency in Eq.(\ref{eigen-1}).

To summarize the above discussions, the ``renormalized bands'' or the ``(would-be) quasiparticle'' approach prefers the self-consistent Eq.(\ref{eigen-2}), while the zero frequency Green's function approach in this paper makes use of Eq.(\ref{eigen-1}).
It has been emphasized\cite{wang2012c} that the self-consistent ``renormalized bands'' or ``quasiparticle'' approach following Eq.(\ref{eigen-2}) is not suitable for calculating topological invariants, though it is an accurate tool for obtaining the energy spectra. On the other hand, the seemingly inconsistent zero frequency Green's function approach is an exact tool for topological invariants\cite{wang2012c}, though it is a poor tool for the purpose of energy spectra.

\section{Strongly correlated topological phases transitions via generalized Chern number}\label{sec:application}

Since our new formalism of topological order parameter can be applied to general interacting superconductors, we will use a simple ansatz of Green's function to illustrate its usage. Suppose that the Green's function of a 2d superconductor is given as \bea \mathcal{G}^{-1}(i\omega,k) = \frac{i\omega+ n\cdot \tau}{ (\omega^2+n^2)^{\gamma}} \label{ansatz} \eea
where $n=(n_x,n_y,n_z)=(\sin k_x,\sin k_y,m+2-\cos k_x -\cos k_y)$, and $\tau=(\tau_x,\tau_y,\tau_z)$ are the Pauli matrices.  
In the $k\rightarrow 0$ limit, this ansatz becomes $\mathcal{G}^{-1}(i\omega,k) \approx (i\omega+ k\cdot \tau +m\tau_z) (\omega^2+k^2 +m^2)^{-\gamma}$.
If $\gamma=0$, this ansatz describes a mean field superconductor with essentially free BdG quasi-particles. Generally, this ansatz describes a strongly correlated superconductor. When we replace $m\tau_z$ term by a $k_z\tau_z$ term, this Green's function becomes the ``unparticle'' propagator\cite{georgi2007,volovik2011}, with power law Green's function which is characteristic of conformal field theories. Unlike these gapless systems, our Green's function ansatz Eq.(\ref{ansatz}) describes a gapped system (for fermion), and the topological order parameter can be readily calculated using Eq.(\ref{chern}).

Following the formalism in the previous section, we write down the zero frequency Green's function as
\bea \mathcal{G}^{-1}(0,k) = \frac{n\cdot \tau} {(n^2)^{\gamma}} \eea  The Berry curvature $\mathcal{F}_{xy}$ can be obtained as \bea \mathcal{F}_{xy}=\frac{1}{2}\epsilon^{abc} \hat{n}_a \partial_{k_x} \hat{n}_b \partial_{k_y} \hat{n}_c \label{berry} \eea
where $\hat{n}^a = n^a/\sqrt{n_1^2+n_2^2+n_3^2}$.
It follows from Eq.(\ref{chern}) and Eq.(\ref{berry}) that \bea C_1(m)=\left\{\begin{array}{cc} 0,&m>0 \\ -1,& -2<m<0 \\1,&-4<m<-2 \\0,&m<-4 \label{ansatz-chern}\end{array}.\right.\eea
It is very interesting to note that the highly nontrivial power $\gamma$ does not enter the topological number because only the Green's function eigenvectors are needed, which do not depend on $\gamma$.   The simplest way to obtain Eq.(\ref{ansatz-chern}) is outlined as follows. The generalized Chern number given by Eq.(\ref{chern}) is unchanged as we tune $m$ except at the three points $m=0,-2,-4$, where the Green's function encounters singularities. Near these three singular points, we can linearly expand $\hat{n}^a$, and obtain the changes of $C_1$. For instance, $\Delta C_1 (m=0) = C_1(m\rightarrow 0^+) - C_1(m\rightarrow 0^-) = -1$ can be obtained by linearly expansion $n\approx (k_x,k_y,m)$ near $k=0$.

The central point here is that the ansatz in Eq.(\ref{ansatz}) describes a strongly correlated superconductor, which can nevertheless be detected by the generalized Chern number for interacting superconductors in Eq.(\ref{chern}). At the topological phase transition point, e.g. when $m=0$, the situation becomes more interesting. Green's function becomes singular at this point, but it is does not fit into the non-interacting picture when $\gamma\neq 0$, since the small $k$ behavior of Green's function is \bea \mathcal{G}(i\omega,k) \sim \frac{(\omega^2+k^2)^\gamma}{i\omega-k\cdot\tau} \eea which signifies a strongly correlated topological phase transition. If $\gamma>1/2$, $\mathcal{G}(i\omega,k)\rightarrow 0$ when $(i\omega,k)\rightarrow(0,0)$, in sharp contrast to the non-interacting transitions with $\mathcal{G}(i\omega,k)\rightarrow \infty$.

It is worth mentioning that the same ansatz as Eq.(\ref{ansatz}) can be proposed for the quantum Hall effect, and an analogous strongly correlated topological phase transition can be obtained using the generalized TKNN invariant obtained in Ref.\cite{wang2012a}

\section{Correlated topological superconductors in 3d}\label{sec:3d}

In this section we focus on time reversal invariant superconductors in 3d, which in the non-interacting limit is classified by an integer\cite{schnyder2008,qi2009b}.
In the Appendix \ref{sec:identity}, we show that the Green's function for superconductors at $(i\omega,k)$ and $(-i\omega,-k)$ is intrinsically correlated as illustrated by Eq.(\ref{phs}).  Time reversal symmetry provides an additional symmetry of Green's function relating $(i\omega,k)$ and $(i\omega,-k)$ as illustrated by Eq.(\ref{gt}). From these two equations we can obtain \bea \Sigma \mathcal{G}(i\omega,k) \Sigma^{-1} = - \mathcal{G}(-i\omega,k) \label{chiral} \eea where $\Sigma=-i\mathcal{C}^{-1}\mathcal{T}=\tau_x\otimes \sigma_y$, in which $\tau$ and $\sigma$ are Pauli matrices in particle-hole and spin spaces [see the Appendix \ref{sec:t}]. Since $\Sigma^2=1$, the eigenvalues of $\Sigma$ is $\pm 1$, and we can choose a basis so that $\Sigma$ is diagonal \bea \Sigma =\left( \begin{array}{cc}
1_{N\times N} & \\
& -1_{N\times N} \\
\end{array} \right)
\eea
Due to Eq.(\ref{chiral}), zero frequency Green's function in this basis is given as \bea \mathcal{G}(i\omega=0,k) = \left( \begin{array}{cc}
 & \mathcal{Q}(k) \\
\mathcal{Q}^\dag(k) & \\
\end{array} \right)
\eea
The topological order order parameter of 3d time reversal invariant topological superconductors is defined as the following winding number \bea W =\frac{1}{24\pi^2} \int d^3k \epsilon^{\mu\nu\rho} {\rm Tr}[(\mathcal{Q}\partial_{\mu}\mathcal{Q}^{-1} ) (\mathcal{Q}\partial_{\nu}\mathcal{Q}^{-1}) (\mathcal{Q}\partial_{\rho}\mathcal{Q}^{-1})] \label{winding} \eea which has similar expression as its non-interacting analog\cite{schnyder2008}, however, Eq.(\ref{winding}) is well defined for superconductors in the presence of electron-electron interaction. As a byproduct,
this shows that the integer classification of 3d time reversal invariant topological superconductors is stable with respect to electron-electron interaction, although various possible fractional states are not included in this description. Therefore, free fermion superconductors in different integer class cannot be smoothly connected even if we add fermion-fermion interaction effect.

In the non-interacting limit, it is straightforward to check that Eq.(\ref{winding}) reduces to the winding number defined in terms of non-interacting Hamiltonian\cite{schnyder2008}.

\section{Gravitational topological field theory and Green's function at zero frequency}\label{sec:gravity}

In this section, we focus on physical responses of 3d time reversal invariant topological superconductors and relate them to the Green's function.
Since in superconductors the $U(1)$ symmetry corresponding to fermion number conservation is spontaneously broken, we cannot get a topological field theory by coupling fermion to $U(1)$ gauge field, instead, we can get a gravitational topological field theory since the space-time metric couples to the energy-momentum tensor. The gravitational topological term obtained in Ref.\cite{wang2011b,ryu2012} from chiral anomaly is given as \bea S_\theta &=& \frac{1}{2}\frac{1}{24}\int \frac{\theta}{8\pi^2} {\rm Tr} R\wedge R \nn \\ &=& \frac{1}{1536\pi^2}\int d^3x dt\epsilon^{\mu\nu\rho\sigma} \theta R^\alpha_{\beta\mu\nu}R^\beta_{\alpha\rho\sigma}\label{GravityTFT}
\eea
where $R^\alpha_{\beta\mu\nu}=\partial_\mu \Gamma^\alpha_{\nu\beta}+\Gamma^\eta_{\nu\beta}\Gamma^\alpha_{\mu\eta}-(\mu\leftrightarrow\nu)$, and the differential form $R=\frac{1}{2}R_{\mu\nu}dx^\mu\wedge dx^\nu$. It is worth noting here that the pre-factor $1/2$ in the first line of Eq.(\ref{GravityTFT}) comes from the fact that the fermions are Majorana instead of complex. The form $\frac{1}{8\pi^2}R\wedge R$ is known as the Pontryagin class in mathematical literatures\cite{nakahara1990}. For general superconductors $\theta= \theta(x,t)$ can be space-time dependent, but for time reversal invariant superconductors, $\theta$ is quantized to be $0$ or $\pi$ mod $2\pi$.

As we mentioned in Sec.\ref{sec:intro}, although the above gravitational topological field theory provides a description of interacting superconductors in this class,  there is no explicit formula (even for noninteracting limit) for the $\theta$ angle from the previous works.
Therefore, the remaining problem is to evaluate $\theta$ in Eq.(\ref{GravityTFT}) explicitly. A natural guess is that $\theta$ is proportional to $W$ given in Eq.(\ref{winding}), namely $\theta=W\pi$. However, $\theta$ can only be defined mod $2\pi$ from the bulk information, because a $2\pi$ ambiguity can always be introduced depending on the details of interfaces between superconductors with different $W$. For instance, the simplest surface states of a topological superconductors with $W=2$ can be described by two surface Majorana cones, and the surface effective Hamiltonian reads\cite{qi2009b,wang2011b,chung2009} \bea H_{2D}=\sum_{\alpha=1,2}\sum_k \eta_{\alpha,-k}^T v\left(\sigma_z k_x+\sigma_x k_y\right)\eta_{\alpha,k} \eea where $\eta_{\alpha,k}=(\eta_{\alpha,k\uparrow},\eta_{\alpha,k\downarrow})^T$ is a two component Majorana fermion operator satisfying $\eta_k=\eta_{-k}^\dag$. A mass term of the form $\sum_\alpha m_\alpha \eta^T_{\alpha}\sigma_y\eta_\alpha$ can induce a gap for the surface states. If $m_1,m_2>0$, the surface thermal Hall conductance\cite{wang2011b} is $\kappa_{xy}= 2\times \frac{\pi k_B^2 T}{24\hbar}$; on the other hand, if $m_1>0,m_2<0$, we have $\kappa_{xy}= 0\times \frac{\pi k_B^2 T}{24\hbar}$. Therefore, the bulk topological invariant cannot fully determine the surface thermal Hall conductance. Following the relation\cite{wang2011b,ryu2012} between surface thermal Hall responses and gravitational topological field theory, namely that $\kappa_{xy}= \frac{\theta k_B^2 T}{24\hbar}$ (assuming that the vacuum has $\theta=0$),  we conclude that $\theta=2\pi$ for the case $m_1,m_2>0$, and $\theta=0$ for the case $m_1>0,m_2<0$.  To summarize the above calculations, if we know from the bulk that the topological superconductor has $W=2$, we can at most conclude that $\theta=2\pi$ mod $2\pi$, in other word,  $\theta=0$ is as equally possible as $\theta=2\pi$ for this topological superconductor. Therefore, we should modify our previous guess $\theta=W\pi$ to \bea \theta = W \pi \,\,\,{\rm mod}\,2\pi \label{theta-winding}\eea We will obtain this result by explicit calculation in Appendix \ref{sec:theta}. The Eq.(\ref{theta-winding}) is among the central results of this paper.

Now let us find a different approach to calculate the gravitational $\theta$ angle.
By analogy with the electromagnetic responses in topological insulators\cite{qi2008}, we propose the following expression for the $\theta$ angle
\bea
\theta =&& \frac{1}{240\pi^2} \int_{-\pi}^{\pi}dk_0 d^3k\int_0^\pi dk_4 \textrm{Tr} [ \epsilon^{\mu \nu \rho \sigma \tau}
\mathcal{G}\partial_{\mu}\mathcal{G}^{-1}   \mathcal{G}\partial_{\nu}\mathcal{G}^{-1} \nonumber
\\  &&  \times  \mathcal{G}\partial_{\rho}\mathcal{G}^{-1} \mathcal{G}\partial_{\sigma}\mathcal{G}^{-1}
\mathcal{G}\partial_{\tau}\mathcal{G}^{-1}] \label{gravitygreen} \eea where $k_0=i\omega$ is the continuous Matsubara frequency( in the zero temperature limit), $k_1,k_2,k_3$ are spatial momenta, and $k_4$ is a dimensional extension parameter (Wess-Zumino-Witten parameter). The reference function $\mathcal{G}(k_0,k_1,k_2,k_3; k_4=\pi)$ is chosen as a trivial ``flat-band'' Green's function analogous to that in Ref.\cite{wang2010b} [see Appendix \ref{sec:theta} for explicit form of  reference function for superconductors].

A direct Feynman diagram calculation of gravitational $\theta$ angle is much more involved than the electromagnetic $\theta$ angle, therefore we will justify Eq.(\ref{gravitygreen}) as follows. In the non-interacting limit, Eq.(\ref{gravitygreen}) reduces to the Chern-Simons term
\bea
\frac{\theta}{2\pi} &=& \frac{1}{16\pi^{2}}  \int
d^{3}k\epsilon^{ijk} \textrm{Tr}\{(F_{ij}(k)-\frac{1}{3}i [ A_{i}(k), A_{j}(k)])\cdot A_{k}(k)\} \nn \\ &=& \frac{1}{8\pi^2} \int d^3k \epsilon^{ijk}{\rm Tr}[\partial_i A_j  +\frac{2}{3}i A_i A_j ] A_k \label{noninteracting-cs} \eea
where $A_i^{\alpha\beta}= -i\langle \psi^\alpha |\partial_{k_i}|\psi^\beta\rangle$ are Berry connections defined in terms of Bloch states $|\psi^\alpha\rangle$. For a non-interacting Dirac fermion with Lagrangian $\bar{\psi} (i\gamma^\mu \partial_\mu +m)\psi$\cite{wang2011b,ryu2012}, from Eq.(\ref{noninteracting-cs}) it follows that \bea \frac{\theta(m>0)-\theta(m<0)}{2\pi} =   \frac{1}{2} \eea which is the same as the result obtained from chiral anomaly\cite{wang2011b,ryu2012}. Since non-interacting Hamiltonian can always be deformed to Dirac type at low energy, we conclude that Eq.(\ref{gravitygreen}) is exactly the gravitational $\theta$ angle in the non-interacting limit. For general interacting superconductors, the quantization of $\theta$ (due to time reversal symmetry) dictates that Eq.(\ref{gravitygreen}) is the correct $\theta$ angle, which is analogous to the case of electromagnetic $\theta$ angle\cite{qi2008, wang2010b}.
The $\theta$ value in Eq.(\ref{gravitygreen}) is of ${\rm Z}_2$ character because of the $2\pi$ ambiguity of the dimensional extension\cite{wang2010b}. This is consistent with the fact that the bulk of a superconductor can only determine the surface thermal Hall coefficient mod even integer\cite{wang2011b}, as we have discussed early in this section.

Since we have justified Eq.(\ref{gravitygreen}) as the correct gravitational $\theta$ angle, we can now derive  Eq.(\ref{theta-winding}) from Eq.(\ref{gravitygreen}). The details of this calculation is left to Appendix \ref{sec:theta}.


\section{Generalization to Green's function singularities in gapless systems}\label{sec:gapless}

It is also straightforward to generalize the zero frequency Green's function topological invariants in this paper to gapless systems, e.g. systems with Weyl points in momentum space. The zero frequency topological invariants are easier to use than the topological invariants with frequency integral\cite{volovik2003}. For instance, a zero frequency Green's function topological invariant can be defined for a singular point (Fermi point, or Weyl point) in 3d gapless systems (insulator or superconductor) as

\bea  C_1 &=& \frac{1}{2\pi} \int_S \mathcal{F}  \nn \\ &=& \frac{1}{4\pi}\int dS^i\epsilon^{ijk}  \mathcal{F}_{jk}
\label{chern-gapless} \eea
where we have chosen a small 2d sphere $S$ around this singular point, and $S^i$ is its area element.  The connection and curvature is defined as   $\mathcal{A}_i = -i\sum_{R-space}\langle k\alpha |\partial_{k_i}|k\alpha\rangle$ and $\mathcal{F}_{ij} =\partial_i \mathcal{A}_j - \partial_j \mathcal{A}_i$, where $|k\alpha\rangle$ are orthonormal vectors spanning the R-space.
Equation (\ref{chern-gapless}) is a direct generalization of Eq.(\ref{chern}) to gapless systems.  The construction of a 2d sphere around the singular point can be found in Ref.\cite{volovik2003}, in which topological invariants for Fermi points are defined in terms of frequency integral. Equation (\ref{chern-gapless}) has the advantage that no frequency integral is needed. For the definition of Eq.(\ref{chern-gapless}), it is crucial to realize that the zero frequency Green's function is a Hermitian matrix, so that a Chern number can be defined on the $k$ space.

\section{Conclusions and Discussions}\label{sec:conclusion}

In this paper we have obtained several topological order parameters for interacting superconductors. They are expressed in terms of zero frequency Green's function, and their equivalence to topological order parameters defined with frequency integral is explicitly shown. These new topological order parameters are much easier to calculate in numerical and analytical calculation. They produce topological invariants different from the ``quasiparticle'' or ``renormalized band'' approach, in which Green's function (or self energy ) at nonzero frequency should be self-consistently used.  Taking advantage of the zero frequency Green's function approach, we analyze a nontrivial topological quantum phase transition beyond the non-interacting picture. The topological field theory coefficient, namely the gravitational $\theta$ angle, has been expressed in terms of the zero frequency Green's function.

For numerical calculation with finite size, at first sight the anomalous Green's function $G_B,G_C$ would vanish because of the fermion number mismatch. One viable approach to circumvent this difficulty is to study the off-diagonal long-range order\cite{penrose1956,yang1962}. The idea can be outlined as follows. Since $\langle 0|c_\alpha(\br_1)c_\beta(\br_2)|0\rangle =0 $ (where $|0\rangle$ is the ground state) for any large but finite system, we can calculate $\langle 0|c_\alpha(\br_1)c_\beta(\br_2)c_\gamma^\dag(\br_3)c_\delta^\dag(\br_4)|0\rangle $ instead. In the limit $|\br_1-\br_2|, |\br_3-\br_4| << |\br_1-\br_3|$, this expectation value can be formally decomposed as $\langle c_\alpha(\br_1)c_\beta(\br_2)\rangle \langle c_\gamma^\dag(\br_3)c_\delta^\dag(\br_4)\rangle $, from which $\langle c_\alpha(\br_1)c_\beta(\br_2)\rangle$ can be extracted. In fact, this can be taken as the definition of $\langle c_\alpha(\br_1)c_\beta(\br_2)\rangle$. The only ambiguity is the global phase factor of Cooper pairing, which is not significant for our purpose.

In the above simplified analysis we omit the $\tau$ ( or $\omega$ ) variable, and  this variable can be straightforwardly added as $\langle 0|c_\alpha(\br_1,\tau_1)c_\beta(\br_2,\tau_2)c_\gamma^\dag(\br_3,\tau_3)
c_\delta^\dag(\br_4,\tau_4)|0\rangle$ (or $\langle 0|c_\alpha(\br_1,i\omega_1)c_\beta(\br_2,i\omega_2)c_\gamma^\dag(\br_3,i\omega_3)
c_\delta^\dag(\br_4,i\omega_4)|0\rangle$ ), thus we can extract $\langle c_\alpha(\br_1,\tau_1)c_\beta(\br_2,\tau_2)\rangle$ (or $\langle c_\alpha(\br_1,i\omega_1)c_\beta(\br_2,i\omega_2)\rangle$ ),
whose Fourier transformation into the momentum space leads to the single particle Green's function $G_B$ and $G_C$ needed to define topological invariants.

This is analogous to calculating magnetization in finite system. The naive expectation value of magnetization $\langle{\bf m}(\br)\rangle=0$, but we can calculate $\langle m_a (\br_1) m_b(\br_2)\rangle$ ($a,b=x,y,z$). In the limit when $|\br_1-\br_2|$ is large, we have the formal decomposition $\langle m_a (\br_1) m_b(\br_2)\rangle \approx \langle m_a (\br_1)\rangle\langle m_b(\br_2)\rangle$, where $\langle m_a (\br)\rangle$ can be regarded as the definition of ``magnetization'' in large but finite systems.  This definition is natural because  $\langle m_a (\br)\rangle$ becomes the true  magnetization in the thermodynamical limit. The nonzero magnetization in the thermodynamics remains visible as the long range correlations in large but finite systems, though the latter do not have magnetization in a strict sense.

In all the calculations in this paper, we have assumed that there is no ground state degeneracy (besides the phase of superconducting order parameter). For fractional phases which violate this condition, our formulas are not directly applicable. The extensions of the present approach to such exotic fractional phases will be left for future works.

We conclude with the remark that the zero frequency Green's function approach used in this paper is equally applicable to topological insulators and superconductors in all symmetry classes in the ``periodic table''\cite{kitaev2009}. The recipe is summarized as follows. We can obtain the zero frequency Green's function using analytical or numerical methods, then we can just calculate topological invariants using the ``noninteracting Hamiltonian'' $h(k)=-G^{-1}(\omega=0,k)$ (for insulators) or $h(k) =-\mathcal{G}^{-1}(\omega=0,k)$ (for superconductors).

ZW would like to thank Xiao-Liang Qi and Xi Dai for discussions.  ZW is supported
by Tsinghua University Initiative Scientific Research
Program(No. 20121087986). SCZ is supported by the NSF under grant numbers DMR-0904264 and the Keck Foundation.

\appendix

\section{Exact Green's function identities for superconductors}\label{sec:identity}

In this appendix, by straightforward calculation we will obtain several useful identities of Green's function for superconductors without assuming any additional symmetry such as time reversal symmetry. Variant forms of some of these identities can be found in Ref.\cite{gurarie2011}.  The departure point of our calculation is the Lehmann spectral representation of the four submatrices in $\mathcal{G}(i\omega,k)$ given as \bea (G_A)_{\alpha\beta}(i\omega,k)&=&\sum_{mn}D_{mn}\frac{\langle n|c_{k\alpha}|m\rangle\langle m|c_{k\beta}^\dag|n\rangle}{i\omega-(E_m-E_n)}  \nn \\
(G_B)_{\alpha\beta}(i\omega,k)&=&\sum_{mn}D_{mn}\frac{\langle n|c_{k\alpha}|m\rangle\langle m|c_{-k\beta}|n\rangle}{i\omega-(E_m-E_n)}  \nn \\
(G_C)_{\alpha\beta}(i\omega,k)&=&\sum_{mn}D_{mn}\frac{\langle n|c_{-k\alpha}^\dag|m\rangle\langle  m|c_{k\beta}^\dag|n\rangle}{i\omega-(E_m-E_n)} \nn \\
(G_D)_{\alpha\beta}(i\omega,k)&=&\sum_{mn}D_{mn}\frac{\langle n|c_{-k\alpha}^\dag|m\rangle\langle m|c_{-k\beta}|n\rangle}{i\omega-(E_m-E_n)} \label{lehmann}
\eea
where $|m\rangle$ are exact eigenvectors of $K=H-\mu N$ with eigenvalues $E_m$ ($H$ is the many-body Hamiltonian, $\mu$ is chemical potential, $N$ is particle number),  $D_{mn}=e^{\beta\Omega}(e^{-\beta E_m} +e^{-\beta E_n})$, and $\Omega$ is the thermodynamical potential defined by $e^{-\beta\Omega}={\rm Tr}e^{-\beta(H-\mu N)}$. We do all calculations at finite temperature and then take the zero temperature limit, which is always implied in this paper. There is a important point hidden in the simplified notation in the above equations, namely that in the ``anomalous propagators'' $G_B$ and $G_C$, the naive expressions such as $\langle n|c_{k\alpha}|m\rangle\langle m|c_{-k\beta}|n\rangle$ vanish identically due to mismatch of fermion number, therefore, the precise meaning of the above formal Lehmann representations for $G_B$ and $G_C$ should be understood as follows. Let us focus on the zero temperature limit $\beta\rightarrow\infty$, then we have the naive expression
\bea
(G_B)_{\alpha\beta}(i\omega,k)&=&\sum_m \frac{\langle 0|c_{k\alpha}|m\rangle\langle m|c_{-k\beta}|0\rangle}{i\omega-(E_m-E_0)}  \nn \\ &+&
\sum_n \frac{\langle 0|c_{-k\beta}|n\rangle\langle n|c_{k\alpha}|0\rangle}{i\omega+(E_n-E_0)} \eea where $|0\rangle$ is the ground state.
This expression of $G_B$ vanishes due to fermion number mismatch.
The correct spectral representation should be modified as
\bea
(G_B)_{\alpha\beta}(i\omega,k)&=&\sum_m \frac{\langle 0_{N-2}|c_{k\alpha}|m\rangle\langle m|c_{-k\beta}|0_N\rangle}{i\omega-(E_m-E_0)}  \nn \\ &+&
\sum_n \frac{\langle 0_{N-2}|c_{-k\beta}|n\rangle\langle n|c_{k\alpha}|0_N\rangle}{i\omega+(E_n-E_0)}  \label{correct} \eea
where $|0_N\rangle$ is the ground states with fermion number $N$. Note that two ground states with fermions number $N$ and $N-2$ enter this spectral representation.  The ground state energy ( $H-\mu N$) satisfies $E_0(N)\approx E_0(N-2)$ in the thermodynamical limit, which is simply denoted as $E_0$. Note that we have absorbed the chemical potential into the definition of Hamiltonian, in other word, $E_n$ is the eigenvalue of $H-\mu N$. (If we define $E_n$ as eigenvalues of $H$, we would have $E_0(N)-E_0(N-2)\approx 2\mu$\cite{abrikosov1975}).
Eq.(\ref{correct}) should always be implied wherever the formal spectral representation given in Eq.(\ref{lehmann}) for $G_B$ appears. Similarly modification of $G_C$ is also understood in this paper.

With these preparations of Lehmann representation, we can obtain exact Green's function identities.
The first exact identity in this appendix is given as \bea G_D(i\omega,k)=-G_A^T(-i\omega,-k) \label{ADequation} \eea which can be obtained from the following calculation \bea (G_D)_{\alpha\beta}(i\omega,k) &=&\sum_{mn}D_{mn}\frac{\langle n|c_{-k\alpha}^\dag|m\rangle\langle m|c_{-k\beta}|n\rangle}{i\omega-E_{mn}} \nn \\ &=& \sum_{mn}D_{mn}\frac{\langle m|c_{-k\beta}|n\rangle\langle n|c_{-k\alpha}^\dag|m\rangle}{i\omega-E_{mn}} \nn \\
&=& -\sum_{mn} D_{mn}\frac{\langle m|c_{-k\beta}|n\rangle\langle n|c_{-k\alpha}^\dag|m\rangle}{-i\omega-E_{nm}} \nn \\ &=& -(G_A)_{\beta\alpha}(-i\omega,-k) \nn \eea where we have defined $E_{mn}=E_m-E_n$. We mention that an variant form of Eq.(\ref{ADequation}) has been given in Ref.\cite{gurarie2011}.

The second identity is \bea G_B(i\omega,k)=-G_B^T(-i\omega,-k) \label{gblehmann} \eea which can be calculated as follows \bea  (G_B)_{\alpha\beta}(i\omega,k) &=&\sum_{mn}D_{mn}\frac{\langle n|c_{k\alpha}|m\rangle\langle m|c_{-k\beta}|n\rangle}{i\omega-E_{mn}} \nn \\  &=&-\sum_{mn}D_{mn}\frac{\langle m|c_{-k\beta}|n\rangle\langle n|c_{k\alpha}|m\rangle}{-i\omega-E_{mn}} \nn  \\  &=& -(G_B)_{\beta\alpha}(-i\omega,-k) \nn \eea
We can check that Eq.(\ref{gblehmann}) can also be obtained directly from the more precise spectral representation given in Eq.(\ref{correct}).

By analogous calculation we can also obtain \bea G_C(i\omega,k)=-G^T_C(-i\omega,-k) \eea

Now we define a charge conjugation matrix \bea \mathcal{C}= \left( \begin{array}{cc} & 1 \\
                  1 & \\
                  \end{array} \right) \nn
                  \eea
then we can readily obtain that \bea \mathcal{C}\mathcal{G}(i\omega,k)\mathcal{C}^{-1} = -\mathcal{G}^T (-i\omega,-k) \label{phs} \eea which can be calculated as \bea \mathcal{C}\mathcal{G}(i\omega,k)\mathcal{C}^{-1} &=& \left( \begin{array}{cc}
G_D(i\omega,k) & G_C(i\omega,k) \\
G_B(i\omega,k) & G_A(i\omega,k) \\
\end{array} \right) \nn \\
&=& \left( \begin{array}{cc}
-G_A^T(-i\omega,-k) & -G_C^T(-i\omega,-k) \\
-G_B^T(-i\omega,-k) & -G_D^T(-i\omega,-k) \\
\end{array} \right) \nn \\
&=& -\mathcal{G}^T(-i\omega,-k) \nn
\eea

By similar calculations, we can also obtain that \bea G_B(i\omega,k) =G_C^\dag (-i\omega,k) \eea
and
\bea G_A(i\omega,k) =G_A^\dag (-i\omega,k) \eea
from which it follows that \bea \mathcal{G}(i\omega,k)=\mathcal{G}^\dag (-i\omega,k) \label{hermitian} \eea

\section{Derivation of $N_2=C_1$ for superconductors}\label{sec:derivation}

The key idea of this derivation is analogous to the one given in Ref.\cite{wang2012a}, though the calculations for superconductors is a little more involved because of appearance of four submatrices $G_{A},G_B,G_C, G_D$ in the Green's function $\mathcal{G}(i\omega,k)$, as we will show below.

The key idea is to introduce a deformation of $\mathcal{G}$ which smoothly connects $\mathcal{G}(i\omega,k)$ and an ``effective non-interacting'' Green's function $(i\omega+\mathcal{G}(0,k)^{-1})^{-1}$ as \bea  \mathcal{G}(i\omega,k,\lambda)=(1-\lambda) \mathcal{G}(i\omega,k)+\lambda (i\omega+\mathcal{G}^{-1}(0,k))^{-1} \label{deformation} \eea where $\lambda\in[0,1]$. It is not evident that this deformation is smooth, since in principle $\mathcal{G}(i\omega,k,\lambda)$ can have zero eigenvalues. It turns out that zero eigenvalue does not appear because of intrinsic properties of Green's function as we show below. It is convenient for our purpose to decompose $\mathcal{G}=\mathcal{G}_1+i\mathcal{G}_2$, where $\mathcal{G}_1,\mathcal{G}_2$ are both Hermitian matrices. The explicit form of $\mathcal{G}_2$ is given by its four submatrices \bea  (G_{2A})_{\alpha\beta}(i\omega,k)&=&-\sum_{mn}d_{mn}\langle n|c_{k\alpha}|m\rangle\langle m|c_{k\beta}^\dag|n\rangle \nn \\
(G_{2B})_{\alpha\beta}(i\omega,k)&=&-\sum_{mn}d_{mn}\langle n|c_{k\alpha}|m\rangle\langle m|c_{-k\beta}|n\rangle \nn \\
(G_{2C})_{\alpha\beta}(i\omega,k)&=&-\sum_{mn}d_{mn}\langle n|c_{-k\alpha}^\dag|m\rangle\langle m|c_{k\beta}^\dag|n\rangle \nn \\ (G_{2D})_{\alpha\beta}(i\omega,k)&=&-\sum_{mn}d_{mn}\langle n|c_{-k\alpha}^\dag|m\rangle\langle m|c_{-k\beta}|n\rangle \label{img2}
\eea where $d_{mn}=D_{mn}\frac{\omega}{\omega^2+(E_m-E_n)^2}$, and $D_{mn}$ is defined in the Appendix \ref{sec:identity}. The crucial property of $\mathcal{G}_2$ is the following. Taking an arbitrary $2N$ component vector $|\psi\rangle = (|a\rangle,|b\rangle)^T$, in which $|a\rangle,|b\rangle$ are $N$-component vectors, we have \bea \langle \psi|\mathcal{G}_2|\psi\rangle &=& -\sum_{mn}d_{mn}[|A_{mn}|^2 +  A_{mn}^*B_{mn} +\ A_{mn}B_{mn}^* +  |B_{mn}|^2] \nn \\ &=& -\sum_{mn}d_{mn}|A_{mn}+B_{mn}|^2 \label{g2sign} \eea Here we have defined $A_{mn}=\sum_{\alpha} a_\alpha^*\langle n|c_{k\alpha}|m\rangle$ and $B_{mn}=\sum_\alpha b_\alpha^* \langle n|c_{-k\alpha}^\dag|m\rangle$ (here $a_\alpha,b_\alpha$ are the components of $|a\rangle,|b\rangle$). The following calculation is almost identical to the case of insulators\cite{wang2012a}, which we will briefly outline. From Eq.(\ref{g2sign}) it follows that \bea {\rm sign}(\langle \psi|\mathcal{G}_2(i\omega,k)|\psi \rangle) = -{\rm sign}\omega \label{g2sign-1}\eea for an arbitrary $|\psi\rangle$. Suppose that $|\psi\rangle$ is an eigenvector of $\mathcal{G}(i\omega,k,\lambda)$ with eigenvalue $\mu_\psi^{-1}$, then it can be obtained that ${\rm Im}[\mu_\psi^{-1}(i\omega,k,\lambda)]=\langle \psi |\psi\rangle^{-1} [(1-\lambda) \langle \psi |\mathcal{G}_2(i\omega,k)|\psi\rangle -\lambda \omega \sum_s |\psi_s|^2 (\omega^2+\epsilon_s^2)^{-1}]$, where we have expanded $|\psi\rangle =\sum_s\psi_s (i\omega,k,\lambda)|s(k)\rangle$, with $|s(k)\rangle$ being the eigenvector of $-\mathcal{G}^{-1}(0,k)$ with eigenvalue $\epsilon_s$. From this equation and Eq.(\ref{g2sign-1}) we can see that when $i\omega\neq 0$, ${\rm Im}[\mu_\psi^{-1}(i\omega,k,\lambda)]\neq 0$. On the other hand, when $i\omega=0$, $\mathcal{G}(0,k,\lambda)=\mathcal{G}(0,k)$ is independent of $\lambda$, and ${\rm Re}[\mu_\psi^{-1}(i\omega,k,\lambda)] \neq 0$. Summarizing the above calculations, we have $\mu_\psi^{-1}(i\omega,k,\lambda)\neq 0$. Therefore, the deformation given in Eq.(\ref{deformation}) is smooth, and we have $N_2(\lambda=0)=N_2(\lambda=1)$. Now we can calculated $N_2=N_2(\lambda=0)$ by calculating $N_2(\lambda=1)$ with ``effective non-interacting'' Green's function $(i\omega+\mathcal{G}(0,k)^{-1})^{-1}$. By a straightforward calculation, we are led to Eq.(\ref{derivation}).

The interested readers are referred to Ref.\cite{wang2012a} for a similar derivation for the case of insulators.

\section{Exact Green's function identities in the presence of time reversal symmetry}\label{sec:t}

The basic calculational tool in this section is the Lehmann representation and the following identity \bea \langle \bar{a}|b\rangle = (-1)^{N_b}\langle \bar{b}|a\rangle,\,\, \langle a|\bar{b}\rangle = (-1)^{N_a}\langle b|\bar{a}\rangle \label{t} \eea which follows from $\hat{T}^2|a\rangle=(-1)^{N_a}|a\rangle$, where $N_a$ is the fermion number of $|a\rangle$, and $|\bar{a}\rangle =\hat{T}|a\rangle$. It is worth noting that Eq.(\ref{t}) are identities for exact many-body eigenstates $|a\rangle$ and $|b\rangle$, which should not be confused with the analogous identitties for non-interacting Bloch states, in which $(-1)^N$ factor is replaced by $-1$. Taking advantage of time reversal symmetry,
we will obtain several exact identities. The first one is \bea G_A^{T}(i\omega,k)=TG_A(i\omega,-k)T^\dag \label{gat} \eea where $T=i\sigma_y$ is the time reversal symmetry matrix defined by $\hat{T}=i\sigma_y \hat{K}$ ($\hat{K}$ is the complex conjugation operator). The transformation of fermion operator is $\hat{T}c_{k\alpha}\hat{T}^{-1}=\sum_\beta T_{\alpha\beta}c_{-k\beta}$.  Eq.(\ref{gat}) is analogous to the time reversal identity in time reversal invariant insulators\cite{wang2010b,gurarie2011}.
Eq.(\ref{gat}) can be derived explicitly as  \bea (G_A)_{\alpha\beta}(i\omega,k) &=& \sum_{mn} D_{mn} \frac{\langle n|c_{k\alpha}|m\rangle\langle m|c_{k\beta}^\dag|n\rangle}{i\omega-E_{mn}} \nn \\ &=& \sum_{mn} D_{mn} \frac{\langle \bar{n}|c_{k\alpha}|m\rangle\langle m|c_{k\beta}^\dag|\bar{n}\rangle}{i\omega-E_{mn}} \nn \\ &=& \sum_{mn} T_{\alpha\gamma}^* T_{\beta\delta} D_{mn} \frac{\langle n|c_{-k\delta}|\bar{m}\rangle\langle \bar{m}|c_{-k\gamma}^\dag|n\rangle}{i\omega-E_{mn}} \nn \\ &=& T_{\beta\delta}(G_A)_{\delta\gamma}(i\omega,-k)T_{\gamma\alpha}^\dag \nn \\ &=& (TG_A(i\omega,-k)T^\dag)_{\beta\alpha} \nn \eea In the third line, we have used the first one in Eq.(\ref{t}) by taking $|a\rangle =|n\rangle$ and $|b\rangle=c_{k\alpha}|m\rangle$, which leads to $|\bar{b}\rangle=\hat{T}c_{k\alpha}\hat{T}^{-1}\hat{T}|m\rangle = T_{\alpha\gamma}c_{-k\gamma}|\bar{m}\rangle$, and finally $\langle \bar{n}|c_{k\alpha}|m\rangle = (-1)^{N_m+1} T_{\alpha\gamma}^*\langle \bar{m}|c_{-k\gamma}^\dag |n\rangle$. By a similar calculation, we can also obtain \bea  G_D^T(i\omega,k)=T^\dag G_D(i\omega,-k)T \eea

The third identity is \bea G_B^T(i\omega,k)=-T^\dag G_C(i\omega,-k)T^\dag \eea which can be explicitly calculated from \bea (G_B)_{\alpha\beta}(i\omega,k) &=& \sum_{mn}D_{mn} \frac{\langle n|c_{k\alpha}|m\rangle \langle m|c_{-k\beta}|n\rangle}{i\omega-E_{mn}} \nn \\ &=& \sum_{mn}D_{mn} \frac{\langle \bar{n}|c_{k\alpha}|m\rangle \langle m|c_{-k\beta}|\bar{n}\rangle}{i\omega-E_{mn}}  \nn \\ &=& \sum_{mn}D_{mn} \frac{T_{\alpha\gamma}^* T_{\beta\delta}^* \langle \bar{m}|c_{-k\gamma}^\dag|n\rangle \langle n|c_{k\delta}^\dag|\bar{m}\rangle}{i\omega-E_{mn}}
\nn \\ &=& \sum_{mn}D_{mn} \frac{T_{\alpha\gamma}^* T_{\beta\delta}^*  \langle n|c_{k\delta}^\dag|\bar{m}\rangle  \langle \bar{m}|c_{-k\gamma}^\dag|n\rangle }{i\omega-E_{mn}}  \nn \\ &=& -T_{\beta\delta}^\dag (G_C)_{\delta\gamma}(i\omega,-k) T_{\gamma\alpha}^\dag \nn \\ &=& -(T^\dag G_C(i\omega,-k)T^\dag)_{\beta\alpha} \nn \eea in which we have used Eq.(\ref{t}) several times and the identity $T^\dag = -T^*$, which follows from $T^\dag T=1$ and $T^*T=-1$. Similarly we can obtain \bea G_C^T(i\omega,k)=-T G_B(i\omega,-k)T  \eea Now we can define a time reversal matrix in the Nambu space as \bea \mathcal{T}= \left( \begin{array}{cc}
T & \\
  & -T^\dag \\
\end{array} \right) \eea With this definition, we can calculate \bea \mathcal{T}\mathcal{G}(i\omega,k)\mathcal{T}^{-1} &=& \left( \begin{array}{cc}
TG_A(i\omega,k)T^\dag & -TG_B(i\omega,k)T \\
-T^\dag G_C(i\omega,k)T^\dag &  T^\dag G_D(i\omega,k)T \\ \end{array} \right) \nn \\ &=& \left( \begin{array}{cc}
G_A^T(i\omega,-k) & G_C^T(i\omega,-k) \\
G_B^T(i\omega,-k) &  G_D^T(i\omega,-k) \\ \end{array} \right) \nn \\ &=& \mathcal{G}^T(i\omega,-k) \nn \eea which is written compactly as  \bea \mathcal{T}\mathcal{G}(i\omega,k)\mathcal{T}^{-1} = \mathcal{G}^T(i\omega,-k) \label{gt} \eea

\section{Green's function in the Majorana operator basis}\label{sec:majorana}
Since the Majorana basis is convenient for some problems, we will present several identities in this basis for references. Let us start with notations. The fermion operators are referred to as $c_{k\alpha}$, where $k$ is spatial momentum, $\alpha=1,2\dots N$ refers to any other degrees of freedom including spin, orbital, etc. We also define $2N$ Majorana operators $\gamma_{k\beta},\,\beta=1,2\dots 2N$ by   \bea \gamma_{k\alpha}=c_{k\alpha}+c^\dag_{-k\alpha},\,\gamma_{k ,N+\alpha}=i(c_{k\alpha}-c^\dag_{-k\alpha}); \,\,(\alpha=1,2\dots N) \eea

Since our results do not depend on the basis we use, we will formulate them in the most convenient basis, namely the Majorana basis defined above.  The Matsubara Green's function in this basis is defined as \bea \tilde{\mathcal{G}}_{\alpha\beta}(\tau,k)=-\langle T_\tau \gamma_{k\alpha}(\tau)\gamma_{k\beta}(0) \rangle \eea where $T_\tau$ is the imaginary time ordering. The frequency domain Green's function is defined as $ G(i\omega_n,k)=\int_0^{\beta}d\tau \exp(i\omega_n\tau)G(\tau,k)$. The Lehmann representation of Green's function at finite temperature is given by \bea \tilde{\mathcal{G}}_{\alpha\beta}(i\omega,k)=\sum_{mn} e^{\beta\Omega}\frac{\langle n|\gamma_{k\alpha}|m\rangle \langle m|\gamma_{-k\beta}|n\rangle}{i\omega-(E_m-E_n)}(e^{-\beta E_m}+e^{-\beta E_n}) \eea  where $\beta$ is the inverse temperature, $|m\rangle$ are exact eigenvectors of $K=H-\mu N$ ($\mu$ is the chemical potential, $N=\sum_{k\alpha}c^\dag_{k\alpha}c_{k\alpha}$ is the fermion number operator), and $\Omega$ is the thermodynamic potential defined by $e^{-\beta\Omega}={\rm Tr}e^{-\beta(H-\mu N)}$. Taking advantage of this spectral representation, we can obtain several exact identities. The first one is \bea \tilde{\mathcal{G}}^\dag(i\omega,k)=\tilde{\mathcal{G}}(-i\omega,k) \eea which is the same as the identity in insulators. The second one is \bea \tilde{\mathcal{G}}(i\omega,k)=-\tilde{\mathcal{G}}^T(-i\omega,-k)  \eea  which has no counterpart in insulators. This identity signifies major difference between the superconductors and insulators, and will be crucial for construction of interacting topological order parameters for superconductors in 3d. It can be appreciated that in contrast to the case in insulator, superconductors's Green's function at $k$ and $-k$ are intrinsically related because of the cooper pairing between opposite momenta.

In the zero temperature limit, the discrete Matsubara frequency variable $\omega$ becomes continuous, and the Lehmann representation of the Matsubara Green's function reads \bea \tilde{\mathcal{G}}_{\alpha\beta}(i\omega)=\sum_m [\frac{\langle 0|\gamma_{k\alpha}|m\rangle \langle m|\gamma_{-k\beta}|0\rangle } {i\omega-(E_m-E_0)} + \frac{\langle 0|\gamma_{-k\beta}|m\rangle \langle\gamma_{k\alpha}|0\rangle} {i\omega+(E_m-E_0)}] \eea
where $|0\rangle$ is the ground state.

We mention in passing that the derivations in Sec.\ref{sec:derivation}  can be simplified if we use the Majorana basis, in which the complication of four submatrices in Eq.(\ref{img2}) can be avoided.

\section{Gravitational $\theta$ coefficient and zero frequency Green's function}\label{sec:theta}

In this section we derive Eq.(\ref{theta-winding}),  which relates the gravitational $\theta$ coefficient to Eq.(\ref{winding}). First, we extrapolate the Green's function $\mathcal{G}(k_0=i\omega,k_1,k_2,k_3)$ to $\mathcal{G}(k_0,k_1,k_2,k_3,k_4)$, with the reference Green's function at $k_4=\pi$ chosen as $G_A(i\omega,k_1,k_2,k_3,\pi)=\frac{1}{i\omega}1_{N\times N}$, $G_D(i\omega,k_1,k_2,k_3,\pi)=\frac{1}{i\omega}1_{N\times N}$, $G_B(i\omega,k_1,k_2,k_3,\pi)= \frac{1}{\Delta} T^\dag$, $G_C(i\omega,k_1,k_2,k_3,\pi)=\frac{1}{\Delta}T$, where $\Delta$ is a positive number with dimension of energy, and $T=i\sigma_y$ is the time reversal operator in the spin space. Physically this Green's function describes a mean field trivial s-wave superconductor with momentum independent pairing.

By an calculation analogous to that given in Ref.\cite{wang2012a} [see also Appendix \ref{sec:derivation}], we can deform $\mathcal{G}(i\omega,k)$ to $(i\omega+\mathcal{G}^{-1}(0,k))^{-1}$ without encountering singularities. This fact will be crucial for the entire calculation, because it implies that zero frequency Green's function can fully determine the topological invariants.

Taking advantage of the time reversal symmetry Eq.(\ref{gt}), we can also write $\theta$ in Eq.(\ref{gravitygreen}) as
\bea
\theta =&& \frac{1}{2}\frac{1}{240\pi^2} \int_{-\pi}^{\pi}dk_0 d^3k\int_{-\pi}^\pi dk_4 \textrm{Tr} [ \epsilon^{\mu \nu \rho \sigma \tau}
\mathcal{G}\partial_{\mu}\mathcal{G}^{-1}   \mathcal{G}\partial_{\nu}\mathcal{G}^{-1} \nonumber
\\  &&  \times  \mathcal{G}\partial_{\rho}\mathcal{G}^{-1} \mathcal{G}\partial_{\sigma}\mathcal{G}^{-1}
\mathcal{G}\partial_{\tau}\mathcal{G}^{-1}] \label{gravitygreen-1} \eea   in which $\mathcal{G}$ in the $k_4 \in [-\pi,0]$ region is determined  by $\mathcal{T}\mathcal{G}(i\omega,k_1,k_2,k_3,k_4)\mathcal{T}^{-1}
=\mathcal{G}^T(i\omega,-k_1,-k_2,-k_3,-k_4)$ and $\mathcal{C}\mathcal{G}(i\omega,k_1,k_2,k_3,k_4)\mathcal{C}^{-1}
=-\mathcal{G}^T(-i\omega,-k_1,-k_2,-k_3,-k_4)$. Eq.(\ref{gravitygreen-1}) is evidently unchanged in smooth deformations. After deforming the Green's function to $(i\omega+\mathcal{G}^{-1}(0,k))^{-1}$ in a similar fashion to Appendix \ref{sec:derivation}, by a direct calculation we can express $\theta$ in Eq.(\ref{gravitygreen}) in terms of zero frequency Green's function explicitly as
\bea \theta/2\pi &=& \frac{1}{32\pi^2}\int_0^{\pi}dk_4\int_{-\pi}^{\pi} d^3k\epsilon^{ijkl}{\rm
tr}[\mathcal{F}_{ij}\mathcal{F}_{kl}] \nonumber \\ &=& {\rm CS}(k_4=0)-{\rm CS}(k_4=\pi) \nn \\ &=& {\rm CS}(k_4=0) \eea
in which the Chern-Simons term \bea
{\rm CS} &=& \frac{1}{16\pi^{2}}  \int
d^{3}k\epsilon^{ijk} \textrm{Tr}\{(\mathcal{F}_{ij}(k)-\frac{1}{3}i [ \mathcal{A}_{i}(k), \mathcal{A}_{j}(k)])\cdot \mathcal{A}_{k}(k)\} \nn \\ &=& \frac{1}{8\pi^2} \int \epsilon^{ijk}{\rm Tr}[\partial_i\mathcal{A}_j  +\frac{2}{3}i \mathcal{A}_i \mathcal{A}_j ] \mathcal{A}_k \label{cs} \eea
where
\bea  && \mathcal{F}^{\alpha\beta}_{ij}=\partial_i
\mathcal{A}^{\alpha\beta}_j-\partial_j
\mathcal{A}^{\alpha\beta}_i+i\left[\mathcal{A}_i,\mathcal{A}_j\right]^{\alpha\beta},\nonumber\\ &&
\mathcal{A}_i^{\alpha\beta}(k) = -i\langle k \alpha|\frac{\partial}{\partial k_i }|k\beta\rangle\nonumber
\eea
in which $|k\alpha\rangle$ is an orthonormal basis of the R-space spanned by R-zeros [see the main text, below Eq.(\ref{eigen}), or see Ref.\cite{wang2012,wang2012a}] . In the $k$ integral in ${\rm CS}$ term, $k_4$ is fixed to be constant, and $d^3k=dk_1dk_2dk_3$. We have used the fact that ${\rm CS}(k_4=\pi)=0$ because $\mathcal{G}(k_4=\pi)$ is a trivial reference function.

Now we can proceed to calculate the Chern-Simons term, which is at this stage already expressed in terms of zero frequency Green's function. We will show that we are led exactly to Eq.(\ref{theta-winding}) by direct calculation.
The physical Green's function at zero frequency has the following form as obtained in the main text \bea \mathcal{G}(0,k_1,k_2,k_3) = \left( \begin{array}{cc}
 & \mathcal{Q}(k_1,k_2,k_3) \\
\mathcal{Q}^\dag(k_1,k_2,k_3) & \\
\end{array} \right)
\eea or \bea \mathcal{G}^{-1}(0,k_1,k_2,k_3) = \left( \begin{array}{cc}
 & (\mathcal{Q}^\dag(k_1,k_2,k_3) )^{-1}\\
\mathcal{Q}^{-1}(k_1,k_2,k_3) & \\
\end{array} \right)
\eea
in which $k_4=0$ is implied.  From the chiral symmetry $\Sigma \mathcal{G}(0,k)\Sigma^{-1} = - \mathcal{G}(0,k)$ it follows that the eigenvectors $|\alpha\rangle=(u,v)^T$ and $|\beta\rangle=(u,-v)^T$ of $\mathcal{G}^{-1}(0,k)$ form chiral pairs as \bea \mathcal{G}^{-1}(0,k)||\alpha\rangle &=& \mu_\alpha |\alpha\rangle \nn \\ \mathcal{G}^{-1}(0,k)|\beta\rangle &=& \mu_\beta |\beta\rangle \nn \eea with \bea
\mu_\alpha &=& -\mu_\beta \nn \eea All R-zeros (namely eigenvectors $|\alpha(0,k)\rangle$ with $\mu_\alpha(0,k)>0$) span the ``R-space''\cite{wang2012a} at each $k$, on which the Chern-Simons term is defined. Due to the specific form of $\mathcal{G}^{-1}$, we can find all the R-zeros by finding the eigenvector of $\mathcal{Q}^\dag\mathcal{Q}$. Suppose that $\mu^{2}\mathcal{Q}^\dag\mathcal{Q}|v\rangle =|v\rangle$ with $\langle v|v\rangle=1$, then $|\alpha\rangle = (\pm\mu \mathcal{Q}|v\rangle,|v\rangle)^T$ is an eigenvector of $\mathcal{G}^{-1}$ with eigenvalues $\pm \mu$.
Therefore, a basis of R-zeros has the following form \bea |\alpha(0,k)\rangle =\frac{1}{\sqrt{2}} \left(\begin{array}{c}
\mu_\alpha \mathcal{Q}|v_\alpha\rangle \\
|v_\alpha\rangle \\ \end{array} \right) \label{chiralvector-1} \eea with $\mu_\alpha>0$, where $|v_\alpha\rangle$ is an $N$ component column vector satisfying $\mu_\alpha^{2}\mathcal{Q}^\dag\mathcal{Q}|v_\alpha\rangle =|v_\alpha\rangle$ and normalized by $\langle v_\alpha|v_\alpha\rangle=1$ . The magnitude of eigenvalues $\mu_\alpha$ is irrelevant for calculations of topological numbers, so we deform all $\mu_\alpha=m>0$. As a result, the eigenvectors $|\alpha(0,k)\rangle$ in Eq.(\ref{chiralvector-1}) are deformed to  \bea |\alpha(0,k)\rangle  = \frac{1}{\sqrt{2}} \left(\begin{array}{c}
\mathcal{R}|v_\alpha\rangle \\
|v_\alpha\rangle \\ \end{array} \right) \label{chiralvector}\eea
where $\mathcal{R}$ is a dimensionless matrix satisfying $\mathcal{R}^\dag \mathcal{R}=1$ because $\mu_\alpha^{2}\mathcal{Q}^\dag\mathcal{Q}|v_\alpha\rangle =|v_\alpha\rangle$ is preserved during the deformation. To calculate the Chern-Simons term over the R-space in Eq.(\ref{cs}),  we can choose a trivial basis in which $|v_\alpha\rangle$ is $k$-independent (this is possible because the fiber bundle spanned by $|v_\alpha\rangle$ is trivial) , and the R-space Berry connection in this basis is given as \bea \mathcal{A}_i=-\frac{i}{2} \mathcal{R}^{-1} \partial_{k_i}\mathcal{R} \label{gauge} \eea
where the crucial $1/2$ factor comes from the $1/\sqrt{2}$ in Eq.(\ref{chiralvector}), otherwise $\mathcal{A}_i$ would become a pure gauge.   Eq.(\ref{gauge}) is not a pure gauge just because of the crucial $1/2$ factor. This is the key of calculations in this section.

It is worth noting that although Chern-Simons term is basis dependent, a basis change (a ``gauge transformation'') can at most change the Chern-Simons term by an integer. Our central result Eq.(\ref{theta-winding}) is a mod $2\pi$ equation, which is thus not affected by basis choices. The Chern-Simons term in Eq.(\ref{cs}) can be calculated directly from Eq.(\ref{gauge}) as \bea {\rm CS}(k_4=0) &=& \frac{1}{48\pi^2} \int d^3k \epsilon^{\mu\nu\rho} {\rm Tr}[(\mathcal{R}^{-1}\partial_{\mu}\mathcal{R} ) (\mathcal{R}^{-1}\partial_{\nu}\mathcal{R}) (\mathcal{R}^{-1}\partial_{\rho}\mathcal{R})] \nn \\ &=& -\frac{1}{2} W(\mathcal{R}) \eea
which is equal to $-\frac{1}{2} W(\mathcal{Q})$ because smooth deformation cannot change the winding number. The minus sign comes from the exchange of $\mathcal{R}$ and $\mathcal{R}^{-1}$ compared to the definition in Eq.(\ref{winding}).
Summarizing the above calculations,  we can see that \bea \frac{\theta}{2\pi} = {\rm CS}(k_4=0) = \frac{1}{2} W(\mathcal{Q})\,({\rm mod}\,\,{\rm integer}) \eea
which is exactly the Eq.(\ref{theta-winding}).

\bibliography{TSC}

\end{document}